%% file: box-wisec2020.tex
\PassOptionsToPackage{table}{xcolor} 
\PassOptionsToPackage{draft}{hyperref}

\documentclass[sigconf,authorversion]{acmart}

\usepackage[final]{pdfpages}
\usepackage[acronym]{glossaries}
\input{acronyms}
\usepackage{tabularx}
\usepackage{multirow}
\usepackage{multicol}
\usepackage[]{xcolor} 

\newcommand{\etal}{\textit{et al.\,}}	

\newcommand\numberof{31 }
\newcommand\numberofnerds{20 }
\newcommand\numberofothers{11 }

\setcopyright{acmcopyright}

\copyrightyear{2020}
\acmYear{2020}
\setcopyright{acmlicensed}\acmConference[WiSec '20]{13th ACM Conference on Security and Privacy in Wireless and Mobile Networks}{July 8--10, 2020}{Linz (Virtual Event), Austria}
\acmBooktitle{13th ACM Conference on Security and Privacy in Wireless and Mobile Networks (WiSec '20), July 8--10, 2020, Linz (Virtual Event), Austria}
\acmPrice{15.00}
\acmDOI{10.1145/3395351.3399342}
\acmISBN{978-1-4503-8006-5/20/07}

\begin{document}

\title{Secure and User-Friendly Over-the-Air Firmware Distribution in a Portable Faraday Cage}

\author{Martin Striegel}

\orcid{1234-5678-9012}
\author{Johann Heyszl}

\author{Florian Jakobsmeier}

\affiliation{%
  \institution{Fraunhofer AISEC}
  \streetaddress{Lichtenbergstraße 11}
}
\email{martin.striegel@aisec.fraunhofer.de}
\email{johann.heyszl@aisec.fraunhofer.de}
\email{florian.jakobsmeier@aisec.fraunhofer.de}

\author{Yacov Matveev}
\author{Georg Sigl}

\affiliation{%
  \institution{Technical University of Munich}
  \streetaddress{Theresienstraße 90}
}
\email{yacov.matveev@tum.de}
\email{sigl@tum.de}

\renewcommand{\shortauthors}{Striegel, et al.}

\begin{abstract}
Setting up a large-scale \glspl{WSN} is challenging, as firmware must be distributed and trust between sensor nodes and a backend needs to be established.
To perform this task efficiently, we propose an approach named \textit{Box}, which utilizes an intelligent \gls{FC}.
The \gls{FC} acquires firmware images and secret keys from a backend, patches the firmware with the keys and deploys those customized images \gls{OTA} to sensor nodes placed in the \gls{FC}.
\Gls{EM} shielding protects this exchange against passive attackers.
We place few demands on the sensor node, not requiring additional hardware components or firmware customized by the manufacturer.
We describe this novel workflow, implement the Box and a backend system and demonstrate the feasibility of our approach by batch-deploying firmware to multiple \gls{COTS} sensor nodes.
We conduct a user-study with 31 participants with diverse backgrounds and find, that our approach is both faster and more user-friendly than firmware distribution over a wired connection.
\end{abstract}



\maketitle

\input{Introduction2.tex}

\input{Related_Work_New.tex}

\input{Design.tex}

\input{Implementation.tex}
\input{Evaluation.tex}
\input{Conclusion.tex}

\begin{anonsuppress}
	\section{Acknowledgments}
	The authors would like to thank the anonymous reviewers for their valuable feedback.
	Many thanks to Nisha Jacob, Katja Miller, Karolin Striegel, Bodo Selmke and Lukas Auer for discussions and suggestions.
\end{anonsuppress}

\bibliographystyle{ACM-Reference-Format}
\bibliography{OTA_Literature.bib}
\end{document}

%% file: acronyms.tex
\makeglossaries

\newacronym{AP}{AP}{access point}
\newacronym{BAN}{BAN}{body area network}
\newacronym{BT}{BT}{Bluetooth}
\newacronym{CA}{CA}{Certificate Authority}
\newacronym{COTS}{COTS}{commercial off-the-shelf}
\newacronym{DHCP}{DHCP}{Dynamic Host Configuration Protocol}
\newacronym{ECDH}{ECDH}{Elliptic-Curve Diffie-Hellman}
\newacronym{EM}{EM}{electromagnetic}
\newacronym{FC}{FC}{Faraday Cage}
\newacronym{GUI}{GUI}{graphical user interface}
\newacronym{HCI}{HCI}{human-computer interaction}
\newacronym{HSM}{HSM}{hardware security module}
\newacronym{HTTP}{HTTP}{Hypertext Transfer Protocol}
\newacronym{IP}{IP}{Internet Protocol}
\newacronym{IPC}{IPC}{Industrial Computer}
\newacronym{JTAG}{JTAG}{Joint Test Action Group}
\newacronym{LLC}{LLC}{location-limited channel}
\newacronym{MAC}{MAC}{Media-Access-Control}
\newacronym{MCU}{MCU}{microcontroller}
\newacronym{OOB}{OOB}{out of band}
\newacronym{OTA}{OTA}{over-the-air}
\newacronym{PKI}{PKI}{Public Key Infrastructure}
\newacronym{PUF}{PUF}{physically uncloneable function}
\newacronym{RF}{RF}{radio frequency}
\newacronym{ROM}{ROM}{Read-Only Memory}
\newacronym{SBC}{SBC}{single board computer}
\newacronym{SE}{SE}{Shielding Effectiveness}
\newacronym{SOC}{SOC}{System-on-Chip}
\newacronym{SNR}{SNR}{signal to noise ratio}
\newacronym{SUS}{SUS}{System Usability Scale}
\newacronym{SSID}{SSID}{Service Set Identifier}
\newacronym{UART}{UART}{Universal Asynchronous Receive Transmit}
\newacronym{UI}{UI}{user interface}
\newacronym{USRP}{USRP}{Universal Software Radio Peripheral}
\newacronym{USB}{USB}{Universal Serial Bus}
\newacronym{WSN}{WSN}{wireless sensor network}


%% file: Introduction2.tex
\section{Introduction}
\glsresetall
The initial setup phase of a \gls{WSN} usually comprises the \textit{distribution of up-to-date firmware} to the sensor nodes as well as \textit{provisioning}, i.e.~assigning the sensor nodes to a \gls{WSN} and establishing trust between communicating parties, which is typically done by exchanging cryptographic material such as keys or certificates.
To address the latter, various approaches for key distribution have been proposed in literature.
However, those approaches are based on the assumption, that the firmware is "just there", omitting, \textit{how} the distribution of firmware is done.

%
%
Supplying firmware images manually via a programmer and cable to a large number of sensor nodes is tedious and hardly scales.
%
Additionally, when sensors are deployed in harsh environments, they are enclosed in a sealed casing.
This limits access to the physical programming interfaces.
Lastly, the person tasked with setting up the \gls{WSN} will likely not be a security expert, thus human error 
must be expected and dealt with.

Rather than using wired interfaces, we propose a device and work-flow called \textit{Box}, which utilizes \gls{OTA} firmware distribution inside a \gls{FC}.
Sensor nodes are prepared with an \gls{OTA}-capable bootloader by the device manufacturer.
The operator creates sensor node firmware images and individual symmetric keys in his backend.
Images and secret keys are then transferred from the backend to the Box using e.g.,~a wired connection.
Afterwards, the Box can be detached and is portable.
Next,the operator places multiple sensor nodes inside the Box and closes it.
For every sensor node, the Box creates a customized image by patching a unique secret key into the generic image and deploys the customized image \gls{OTA} to a particular node.
%
%
Sensor nodes can immediately begin to read sensors and exchange encrypted and authenticated messages with the backend, as both share a secret key.
%
%
\Gls{EM}-shielding
protects this process against eavesdropping, preventing an attacker from overhearing the secret keys and the firmware, which might be intellectual property.

To the best of our knowledge, we are the first to propose a workflow, which explicitly tackles the challenge of firmware distribution using \gls{OTA} updates in a shielded environment.
We designed and implemented the Box and the backend and demonstrate the complete \gls{OTA} firmware distribution process using \gls{COTS} sensor nodes.
We conducted a user study with \numberof participants to evaluate our approach and found, that the Box surpasses wired distribution in terms of speed and user satisfaction.

\begin{figure*}[!h]
	\centering
	\includegraphics[]{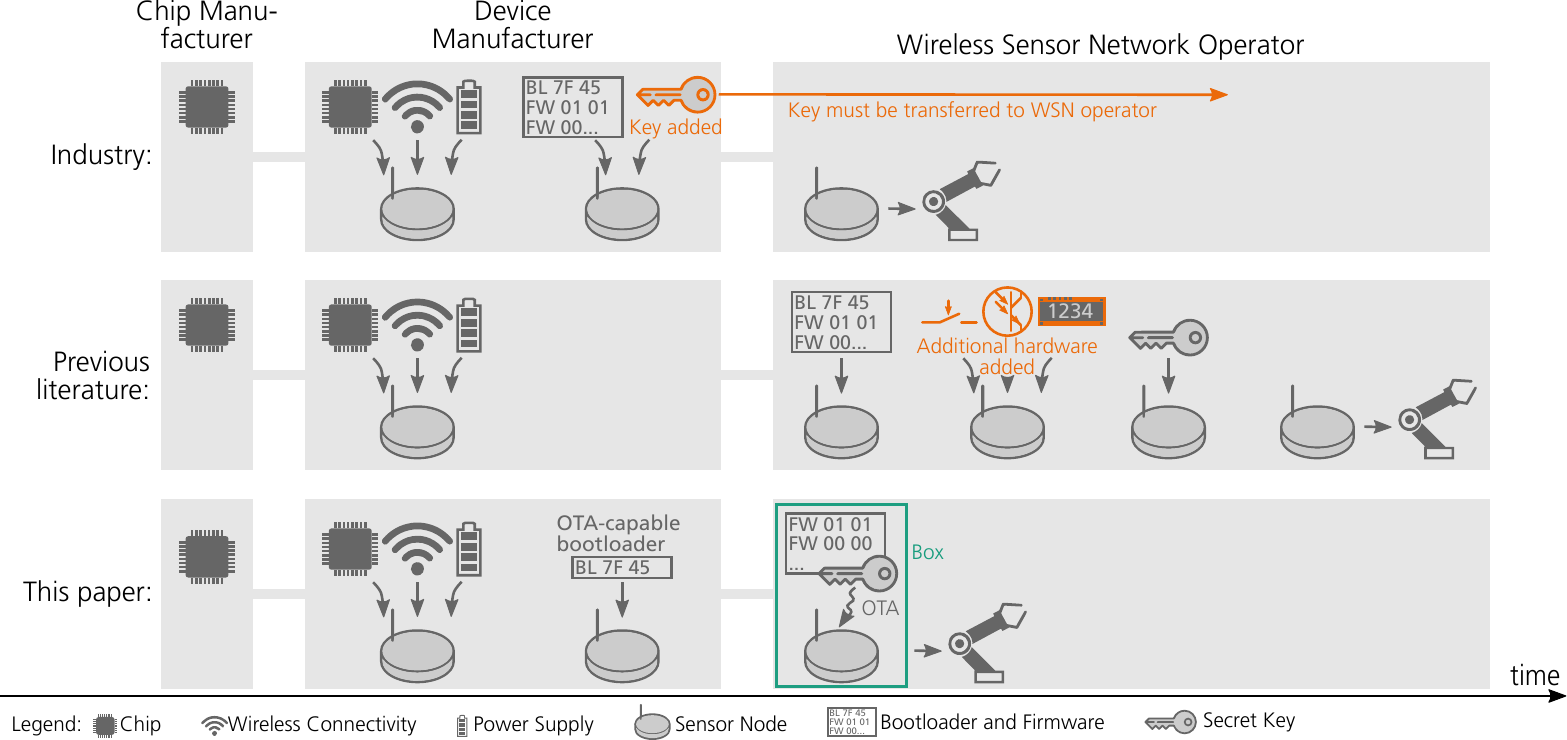}
	\caption{Provisioning options for deploying \glspl{WSN}}
	\label{Fig:SupplyChain}
\end{figure*}

\paragraph{Road-map:} In Section \ref{Sec:RelatedWorkNew}, we review approaches from industry and previous literature.
In Section \ref{Sec:Design}, we describe the scenario, attacker model and key ideas of our approach.
In Section \ref{Sec:Implementation} we outline the implementation.
In Section \ref{Sec:Evaluation}, we conduct a user study and present the results.
Section \ref{Sec:Conclusion} concludes the paper.

%% file: Related_Work_New.tex
\section{Comparison to Other Approaches}
\label{Sec:RelatedWorkNew}
Typically, there are three parties involved in provisioning \glspl{WSN}.
The chip manufacturer produces the microcontroller chips.
The device manufacturer assembles a sensor node by joining chip, sensors, printed circuit board and power supplies.
Lastly, there is the operator, who installs the \gls{WSN}.
Using those role definitions, in Figure \ref{Fig:SupplyChain}, we compare solutions from industry and previous literature to our Box-approach and discuss its benefits.

\paragraph{Industrial Solutions}
In typical industrial solutions, as shown in row one of Figure \ref{Fig:SupplyChain} the burden of supplying firmware images and provisioning is shifted to the device manufacturer.
This means entrusting firmware (which might be confidential intellectual property) as well as the secret key to the device manufacturer.
Subsequently, keys need to be transferred to the \gls{WSN} operator securely.
This approach is only feasible for high volume markets, as device manufacturers might be unwilling to customize individual devices in low volumes.
Lastly, flashing firmware to sensor nodes in an early stage of production forces the \gls{WSN} operator to update the firmware to the most recent version before deployment, causing additional efforts.

\paragraph{Previous Literature}
Approaches from previous literature are shown in row two in Figure \ref{Fig:SupplyChain}.
Those works typically propose to exchange secrets or authentication information using \textit{\glspl{LLC}} as proposed by Stajano and Balfanz \cite{Stajano1999, Balfanz2002}.
\Glspl{LLC} enforce physical proximity and/or directivity.
By doing so, they provide authenticity and, sometimes, confidentiality.
An attacker sending on those channels will be detected by the legitimate operator.
%
%
\Glspl{LLC} often used include visual channels using light \cite{Saxena2006, Saxena2009, Gauger2009, Balfanz2002, Balfanz2004}, sound \cite{Lopes2001, Goodrich2006, Soriente2008} or the operator inserting data \cite{Holmquist2001, Lester2004, Castelluccia2005, Mayrhofer2007}, pointing a provisioning device \cite{Swindells2002, Hoepman2008} or comparing data and confirming equality \cite{Cagalj2006, Prasad2008, Chen2008, Li2010, Perkovic2012, Bluetooth5Specification}.

%

Creating \glspl{LLC} has drawbacks:
Firstly, sensor nodes must be equipped with additional hardware, increasing their cost and size.
Secondly, letting the user perform security-critical tasks, e.g., comparing audio patterns and confirming their equality, is dangerous, as minor operating errors can turn into critical security errors \cite{Cranor2008}.
%

We are aware of three previous works, which propose the use of a \gls{FC}.
The idea to use a \gls{FC} to create a \gls{LLC} has been mentioned in one sentence by Castelluccia and Mutaf \cite{Castelluccia2005}.

Kuo \etal were the first to implement this approach and show, that exchanging symmetric keys between a base station and sensor nodes in a \gls{FC} is very user-friendly \cite{Kuo2007}.
To reduce the \gls{SNR} at the attacker, Kuo \etal use a jammer attached to the exterior of the Faraday cage \cite{Kuo2007}.

Law \etal introduce a key management architecture for body sensor networks, utilizing \gls{ECDH} key exchange inside a \gls{FC} \cite{Law2011}.
They also utilize a jammer attached to the \gls{FC}.
While public key cryptography is beneficial in terms of key management, it creates high demands on the computational performance of a sensor node.
Unlike our approach, their scheme relies on the operator making security-critical decisions by forcing her to configure the \gls{FC}, which is error-prone.

Both Kuo \etal and Law \etal require a jammer outside the \gls{FC}.
While this is beneficial to reduce the \gls{SNR} at the attacker and hence increases the security level, jamming power must adhere to regulations and can not be set to arbitrarily high values.
Further, benign nearby devices might be jammed, too, decreasing their availability.
Our approach does not need a jammer, as we introduce a combination of software and chained hardware attenuators to reduce the transmission power inside the Box to a low level.
This, in addition with the shielding, results in a low \gls{SNR}, effectively preventing eavesdropping by an external attacker.

Kuo \etal and Law \etal assume that a key exchange protocol is present in the firmware of every sensor node.
Their works focus on distributing keys, assuming that the interface and protocol used to exchange keys was delivered to the sensor node a priori by some unspecified mechanism.
Hence, it can be argued, that keys could just have been transferred together with the firmware image over the same wired connection.
We, on the other hand, focus on the \textit{initial distribution} of firmware.

\paragraph{This Paper}
Our Box-approach is shown in row three of Figure \ref{Fig:SupplyChain}.
We require \gls{COTS} microcontrollers to be prepared with a \gls{OTA}-capable bootloader by the chip- or the device manufacturer or a third party.
This is a reasonable assumption, as the \gls{OTA} bootloader need not be customized.
Rather, the Box adapts to the configuration of the bootloader.
Thus, chip or device manufacturers can ship all microcontrollers with this default bootloader.

\gls{OTA} is supported by many commonly used \gls{WSN} transmission protocols and sensor node platforms \cite{BTLE_OTA, WIFI_OTA, CONTIKI_OTA, ZIGBEE_OTA, LORAWAN_OTA}.
The primary technical requirement is a sufficiently fast transmission channel from the \gls{OTA} update server to the sensor node.
Our approach is compatible with all those transmission standards and thus works with many \gls{COTS} sensor nodes.


What are the benefits of our approach?
Users can supply runtime firmware in-house.
This means, that no confidential information such as secret keys, nor the runtime firmware, which is intellectual property of the commissioning party, need to be exposed to a third party.
Providing firmware right before \gls{WSN} deployment guarantees its freshness.
%
%
With the Box, we have a portable secure environment, in which we can update sensor node firmware in the field instead of returning sensor nodes to the facility.
A large number of sensor nodes can be placed in the Box and provisioned simultaneously.
This improves speed and minimizes room for human error, since the number of manual actions performed by the operator is reduced.
Unlike Kuo \etal, who need identical sensor nodes to be able to perform simultaneous key distribution, we can provision different types of sensor nodes at the same time.
%
%
Lastly, we have a clear split between \gls{OTA} bootloader and runtime firmware.
Thereby the runtime firmware can be kept small, which is the main requirement for stable operations and availability \cite{Dawson2012}.

We consider our approach a leap towards real-world applicability, reducing the number of hardware parts needed, not interfering with nearby devices and especially solving the issue of initial firmware distribution, which was previously assumed to happen implicitly.

%% file: Design.tex
\section{Box-Based Solution}
\label{Sec:Design}

  
\paragraph{Overview:}
\label{Subsec:SystemDesign}

\begin{figure}
	\centering
	\includegraphics[width=\columnwidth]{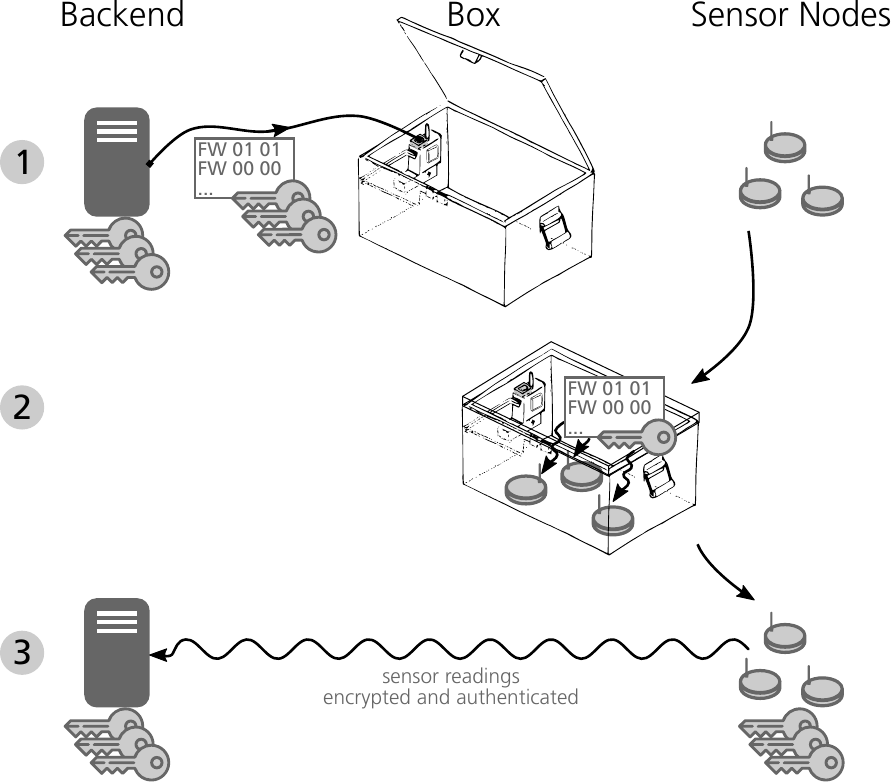}
	\caption{Firmware distribution and provisioning workflow}
	\label{Fig:Workflow}
\end{figure}
Figure \ref{Fig:Workflow} shows the workflow, divided into steps \textcircled{1} to \textcircled{3}.
Our system has three entities: Box, backend and an arbitrary number of sensor nodes.
Initially, the backend creates a firmware image and a number of unique secret keys using a true randomness source.
Next, the operator connects the Box to the backend using e.g.,~a cable (step \textcircled{1}).
This can be considered a proximity-based \gls{LLC}, which achieves security against passive and active attacks.
Additionally, cryptographic mutual authentication between backend and Box should be used.
The Box requests a firmware image and keys from the backend and stores them into its memory.
Now, the Box can be unplugged and is fully portable.

To deploy firmware to sensor nodes, the operator powers them on, places them inside the Box and closes its lid.
Upon detecting, that it is shut tightly, the Box creates a wireless network within.
Each sensor node can now utilize its wireless transceiver to join the network, request and acquire a firmware \gls{OTA} from the Box.
Upon receiving such requests, the Box patches individual keys into the firmware image and sends the now unique firmware to the node.
This process is protected from an external attacker against eavesdropping, as the walls of the box shield \gls{EM} waves and the Box transmits at low power (step \textcircled{2}).

Afterwards, the operator removes the sensor nodes from the Box and deploys them.
Every sensor node reboots, now using the new firmware image.
This image contains functions to read sensors, the address of the backend and credentials to access the network created by the backend.
Now, the sensor node can initiate cryptographically secured communication with the backend, since they share a secret key (step \textcircled{3}).

\paragraph{Attacker Model}
The goal of the attacker is to learn the firmware and secret keys exchanged between the Box and the sensor nodes.
He is computationally bounded, implying that his computing power is polynomial at most.
We assume the attacker to be present before, during and after firmware distribution.
During every phase, he may launch passive or active wireless attacks.

We assume the sensor node supply chain to be trusted including the operator, who deploys the sensor nodes using the Box.
This means, that no desired hardware or software is inserted into a sensor node.
Trusting the supply chain is a common assumption, as searching for hardware implants in every device bought is unfeasible and requires dedicated equipment.

The operator is no security expert, however, he is assumed to be aware of his surroundings, i.e.~able to notice obviously strange events.

We assume that the attacker neither has physical access to the backend, nor to the Box.
He might capture sensor nodes, after they have been set up by the Box.
%
%

\subsection{Box}
\label{Subsec:Box}
\begin{figure*}[]
	\centering
	\includegraphics[width=1\textwidth]{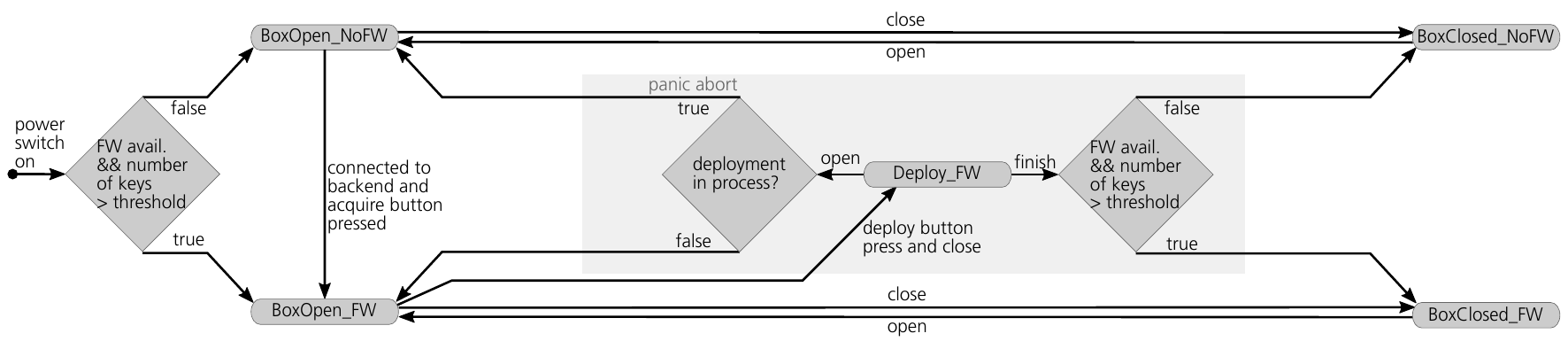}
	\caption{State machine of the Box}
	\label{Fig:StateMachine}
\end{figure*}

The Box comprises a Faraday Cage, a \gls{SBC}, wireless transceivers, battery, an open/close sensor and the \gls{UI}.
The \gls{UI} consists of a speaker, two pushbuttons with built-in status LEDs (the \textit{acquire} and the \textit{deploy} button) and the power button, all placed inside and only accessible, while the Box is open.

The Box is modeled as a five-state state machine, which is shown in Figure \ref{Fig:StateMachine}.
States are based on the Box being opened or closed and whether its internal memory contains firmware images and keys.
Transition between those states is initiated by the operator performing manual actions, such as opening or closing the box and pressing a button.
Security-relevant tasks such as downloading firmware and keys from the backend into the Box and the actual firmware distribution process are conducted automatically by the Box.

The speaker provides information about the current state and what the operator has to do next.
Our study confirmed, that audio is clearly understandable despite the box being closed.
Using a speaker has several benefits over using e.g.,~LEDs:
Firstly, mechanical integrity and \gls{EM} shielding are not impaired, as no wires need to be led through the enclosure of the Box.
Secondly, spoken audio feedback is more explicit than encoding information into LED on/off state and coloring.

The Box utilizes both software and hardware attenuation mechanisms to transmit at very low power barely enough for the sensor nodes inside the Box to receive the messages. 
We can utilize such low transmission power, because the \gls{EM} shielding protects communications inside the Box from external interference, e.g.,~from harsh industrial environments.
Low transmit power in addition to the \gls{EM} shielding protects the firmware-plus-key exchange from eavesdropping attacks by the remote adversary.

\subsubsection{State Machine}
After power-up, the Box checks, if at least one firmware image and a number of keys greater than a preset threshold reside in its memory.
If the actual number is below the threshold, it moves to the \texttt{BoxOpen\_NoFW} state.
The operator connects the Box to the backend and presses the acquire button.
The Box automatically acquires a firmware image and secret keys and stores those.
Then, it moves to the \texttt{BoxOpen\_FW} state.


If the Box is closed, while the deploy pushbutton has not been pressed, the Box transitions into the \texttt{BoxClosed\_FW} state, in which it can be carried conveniently.
The \texttt{Deploy\_FW} state is entered, from the \texttt{BoxOpen\_FW} state, after the operator has pressed the \textit{deploy} pushbutton and closed the box.
Now the Box creates a wireless network using the transmission protocol and connection parameters expected by the sensor nodes.
%
%
For example, if WiFi is used, the name and passphrase of the access point as well as the address, at which the webserver serving the firmware images can be reached, match the factory defaults stored in the \gls{OTA} bootloader of the sensor nodes.
Hence, the Box adapts to the sensor nodes.

If the operator forgot to place nodes inside the Box, the Box waits for requests for a certain time. 
If no requests were received, the Box informs the operator about this.
Then, it moves to the \texttt{BoxClosed\_FW} state, where it resides until it is opened by the operator.

To program sensor nodes, the operator places powered-on sensor nodes with an \gls{OTA} bootloader in the Box, presses the deploy button and closes the Box.
Sensor nodes connect to the wireless network created by the Box, and request firmware images from the box.
Upon receiving an \gls{OTA} request from the sensor nodes, the received keys are patched into the firmware binary and deleted from the key database of the Box.
Binary patching is computationally less expensive then compiling each firmware with its individual key and thus much faster, speeding up the firmware distribution process.
Additionally, the Box does not need development toolchains for sensor nodes, improving maintainability.
It is able to sign firmware images, which is discussed in Section \ref{subsubsec:securitydiscussion}.

The box counts individual device addresses, e.g.,~MAC addresses from which it received requests.
After deployment is finished, the Box informs the operator about the number of sensor nodes provisioned and instructs the operator to open the Box and remove the sensor nodes.
Upon opening, if the remaining number of keys inside the database of the Box is higher than a threshold, the Box moves to the \texttt{BoxOpen\_FW} state, else to the \texttt{BoxOpen\_NoFW} state.

How does the box know, if all sensor nodes have been supplied with firmware?
Our user study described in Section \ref{Sec:Evaluation} shows, that operators are aware of how many nodes they placed inside the box.
Thus, operators notice a mismatch between the number of provisioned nodes the Box announced and the number of nodes they placed inside the Box.
If sensor nodes have an LED, they can issue a blink pattern after having received the new firmware, indicating success.
Else, if an optical marker such as a QR-code is present, solutions such as the one proposed in \cite{Striegel2019} can be used to inspect the sensor node.

While the Box is closed, wireless transmissions inside are secured against eavesdropping by the \gls{EM} shielding in conjunction with the low transmission power.
If the open/close-sensor of the Box detects, that it is opened while deployment is in process, it performs a panic abort, since an opened \gls{FC} does not shield \gls{EM} waves sufficiently.
As the Box expects an attack, it erases all secret keys from its database and informs the operator.
Then, it moves to the \texttt{BoxOpen\_NoFW} state.

\subsubsection{Benefits of Our Approach}
As the pushbuttons and sensors, which cause state transitions, have been placed inside the Box, the mechanical design ensures, that no unintended states are reached, the most dangerous state would be \texttt{Deploy\_FW} while the Box is opened.
This is accomplished by reducing security-critical tasks to simple manual actions.
Those manual actions hide the complexity of \gls{OTA} firmware distribution, while they still allow the operator to determine cause and effect of his actions.
Hence, possibilities for handling errors are minimized by keeping the operator out of the loop while providing her a sense of control \cite{Cranor2008}.

%
%
%

The number of sensor nodes which can be provisioned simultaneously is only limited by the physical dimensions and the maximum number of addresses supported by the network protocol.
Any \gls{COTS} sensor node can be used with the Box, as long as the sensor node uses an \gls{OTA}-capable transmission protocol and has been supplied with an \gls{OTA} bootloader, e.g., by the device manufacturer.
%
%
Our approach does neither require hardware modifications, nor changes to the \gls{OTA} bootloader, as the Box adapts to the configuration of the bootloader and not vice versa.

The Box supports simultaneous distribution of firmware to non-homogeneous sensor nodes.
For example, the operator can place sensor nodes using WiFi and other nodes using Bluetooth in the Box and supply different firmware images to those at the same time.

The Box could also contain multiple firmware images configured for different tasks (e.g., read different sensors mounted at the same sensor node).
Then, the operator could choose \textit{on-site} and right before deployment, which firmware image shall be used and thus which quantity should be metered by the sensor.
%
%
This brings high flexibility: In previous approaches, the decision which firmware is delivered to which node has been made way earlier (e.g., during sensor node production).
Our approach maximizes the time for firmware development, ensuring that the most recent and mature firmware version is on the node.
Finally, the Box can be used to update sensor node firmware in the field.
For this purpose the sensor node is brought into bootloader mode.
Then, the Box is used just like in initial firmware distribution, its shielding providing a portable secure environment.
%

\subsubsection{Shielding for Eavesdropping Security}
\label{SecurityDesign}


%

We want to protect the exchange of firmware with an embedded secret key against a passive remote attacker.
We assume a strong attacker, who knows how the wireless packets are structured, i.e.~how the firmware is split into the payload blocks and where to find the key in a particular firmware chunk.

Obtaining information by eavesdropping is impossible if the received power $P_{RX}$ is lower than the sensitivity of the attacker's receiver.
Additionally, the received signal strength must be below the noise floor, i.e.~the \gls{SNR} is too low to be able to distinguish information from noise.

Using a radio link budget equation and taking into account the attenuation provided by the Box, the signal strength at the attacker can be calculated as:
\begin{equation}
P_{RX} = P_{TX} + G_{TX} - L_{FSPL} + G_{RX} - L_{Box}
\label{Eq:Prx}
\end{equation}
where $P_{RX}$ is the power received by the attacker, $P_{TX}$ the transmission power of the sender, $G_{TX}$ gains at the sender, $G_{RX}$ gains at the receiver, $L_{FSPL}$ the free space path loss and $L_{Box}$ the attenuation of the Box.
We control $P_{TX}$, $G_{TX}$ and $L_{Box}$, while the attacker can influence $L_{FSPL}$ and $G_{RX}$.

The free space path loss is given by:
\begin{equation}
L_{FSPL} = 20\cdot log_{10}(d) + 20\cdot log_{10}(f) - 67.55 
\label{Eq:Lfs}
\end{equation}
where \textit{d} is the distance between sender and receiver in centimeters and \textit{f} the frequency in megahertz.
In Equation \ref{Eq:Prx}, we do not consider fading or losses caused by antenna polarization mismatch. 
Those would \textit{increase} attenuation and contribute to the protection level.
As we calculate a worst-case scenario here, we omit them.

Unlike previous solutions using a \gls{FC}, we do not require a jammer mounted externally to the Box.
First, we determined the minimal transmission power $P_{TX}$, at which the sensor nodes used are still able to receive wireless messages.
For the sensor nodes used in our example, this $P_{TX,min}$ is about $-90\,dBm$ \cite{ESP32Datasheet}.
Hence, we feed signals from the wireless interface of the Box through a chain of four $20\,dB$ hardware attenuators before being passed to the antenna.
We reduce $P_{TX}$ in software to cancel the antenna gain and set $P_{TX}$ so low that sensor nodes within the Box are just able to receive the signal, in our example slightly above $-90\,dBm$.
The \gls{EM} shielding provided by the Box reduces interference from outside the Box, permitting communications at this low transmission power.
As sensor nodes do not transmit confidential information and they must be able, to reach a backend far away, their transmission power does not need to be adjusted.

\paragraph{Example Calculations}
Transmitting at 2.4\,GHz, assuming $L_{Box}$ at $40\,dB$, $L_{FSPL} = 34\,dB$ in a distance of 50\,cm and $G_{RX}$ at $30\,dB$, we calculate $P_{RX,Attacker} = -134\,dBm$.
This lies well below the sensitivity of typical receivers an attacker would use, which is around -96\,dBm \cite{AtherosDatasheet}. 
Note, that we can safely assume that the operator of the Box would notice an attack setup this physically close to him.
The attacker could use more powerful (and thus larger) antennas and wireless receivers to achieve a higher signal strength, however, his presence can be detected more easily then.
To account for that, our hardware-store Box prototype, which attenuates by 40\,dB, can be exchanged for commercial solutions offering shielding greater than $70\,dB$.

The \gls{SNR} at the attacker's receiver is derived as follows:
For exchanging firmware, in our tests, we use 802.11n WiFi at $20\,MHz$ channel width and the maximal data rate of 150\,Mbps supported by the sensor nodes used in our experiments.
Assuming only thermal noise (a worst-case assumption security-wise) at bandwidth $B=20\,MHz$ and room temperature, the noise floor of the attacker's receiver is $-101\,dBm$.
%
The \gls{SNR} is thus $-134\,dBm-(-101\,dBm)=-33\,dB$.
We transfer each individual key only once, so the attacker can not reduce noise by combining multiple transmissions.

To achieve channel capacity for a data rate of $150\,Mbps$, we require a \gls{SNR} of $\frac{S}{N} = 2^{C/B}-1$, which evaluates to $17.4\,dB$.
%
%
Note that this is the best-case assumption for the attacker:
It only holds true using optimal coding, which is not given in practice.

Those calculations show, that even close to the Box and assuming fairly low attenuation by the Box, the received power $P_{RX}$ is lower than the minimal receive power required by a typical attacker.
Further, the \gls{SNR} at the attacker is so low, that channel capacity can not be achieved, hence received information will be erroneous. 
Thus, the attacker will not be able to learn the firmware image and the secret keys by eavesdropping.

To be able to maintain those security levels, the box must be shut tight during firmware distribution, which is checked by the open/close sensor in the \texttt{Deploy\_FW} state.
If the Box is opened during key exchange, all keys are deleted from the memory of the Box and the operator is warned.

\subsubsection{Security Against Active Wireless Attacks}
An attacker might launch a rogue wireless network with the same identifier the Box would use to create its network.
By doing so, sensor nodes are tricked into joining the attacker's network.
There, they are supplied with malicious firmware and network credentials of the attacker's backend network.
Subsequently, sensor nodes would send their sensor readings to the malicious backend network instead of the intended network.

This type of attack is made difficult by the shielding of the Box, which reduces the signal strength of the rogue network inside the Box to low levels.
Thus, sensor nodes inside the closed box will only detect and potentially join that network, if the attacker uses very high transmission power, 
As an additional means of protection, the Box can permanently monitor the wireless spectrum.
If it detects a network with the same identifier and channel setting it would use to create its own wireless network, it issues a spoken warning.
The Box can not prevent, that sensor nodes have joined the rogue network.
The operator is instructed to turn off the sensor nodes and consider them compromised.

\subsubsection{Security Discussion}
\label{subsubsec:securitydiscussion}
How does the backend know, whether a key has been deleted from the box and thus invalidated?
The Box could have a wireless communications channel to the backend, over which those information could be communicated.
As we can not guarantee that this channel is available, the backend also stores the last time a key has been used.
It can be configured to automatically blacklist a key, which has been sent to the Box, but which has not been seen in use by a sensor node for a given time.

How to use signed firmware images with the Box?
The Box can contain a public-private key-pair, whose private key is used for firmware signing.
This key should be stored in a \gls{HSM}.
If a sensor node is placed in the Box for the first time, its \gls{OTA} bootloader does not check the signature of firmware updates.
Adding \textit{user-specific} public keys to the bootloader would imply that the manufacturer adapts the bootloader to each customer, contradicting our approach.
%
%
If the device manufacturer would add a \textit{default} public key to all sensor nodes shipped, the corresponding private key would need to be delivered to all parties acquiring those nodes.
By doing so, the firmware signing process would be rendered useless.
Instead, in our approach during the \textit{very first firmware update}, the Box does not distribute a runtime firmware.
Rather, it delivers a new \gls{OTA} bootloader to the sensor nodes, which now contains the public key of the Box.
This bootloader replaces the old one by overwriting the bootloader sectors in flash.
Now, as the public key is present, all subsequent \gls{OTA} firmware updates must be signed by the Box and the sensor node can check the authenticity of those updates.
Since an remote attacker does not possess the private key needed to sign firmware images and given that debug interfaces are locked, the \gls{OTA} bootloader can not be overwritten again.
Hence, the sensor node now only accepts firmware updates signed by the Box.

How to detect a malicious node inside the Box?
Note, that the Box can not detect hardware implants in sensor nodes, as this would require sophisticated equipment.
For this reason, as stated in the section \textit{Attacker Model}, we must assume the sensor node supply chain including the operator to be trusted. 
If we deviate from this assumption, it could be imagined that an attacker prepares a sensor node with a malicious firmware, which eavesdrops and records all firmware transfers inside the Box.
As the firmware images contain secret keys, the attacker could remove the malicious node later or exfiltrate the confidential data captured via a wireless channel.
This data can than be analyzed to extract all keys distributed in this run of the Box. 
%

In order to prevent this kind of attack from inside the Box, \textit{Secure Erasure} techniques proposed by Perito \etal can be applied \cite{perito2010secure}. 
Those are used after we have updated the \gls{OTA} bootloader with the one containing our public key and before deploying our runtime firmware.
Sensor nodes are forced to delete their memory, thus removing potentially malicious code. 
The malicious node, placed inside the box can act by either answering to the erasure request or ignore it.
A node is not able to fake the verification of the erasure, therefore it is detectable if the memory was cleared.

If all nodes respond, the Box can verify that the whole memory is securely erased, and proceed with firmware and key exchange.
%
%
However, if a node does not communicate at all, the Box can not notice this.
As discussed earlier, it is a reasonable assumption, that, the operator knows, how many nodes were put in the box.
After the firmware update, the Box announces, how many nodes have been securely erased and written with the new firmware. 
The operator intuitively sees the difference of nodes placed into the Box versus nodes supplied with firmware.
Thus, he can detect that not all nodes were flashed. 
He can choose to flash the batch again (e.g~individually or pairwise to reduce the effort and find which nodes were not flashed).
This ensures, that only nodes, which have been erased securely and supplied with the desired firmware image are added to the network.

\subsection{Backend}
\label{Subsec:backend}
The backend has the largest computational resources in the system.
It can utilize a true randomness source and creates secret individual keys.
Additionally, it compiles firmware images for the different sensor node types used.
%
%
Afterwards, the backend waits for the Box to be connected.
Mutual authentication between the backend and the Box can be accomplished using known protocols, e.g., TLS.
Upon receiving a valid download request, the backend transfers firmware images and a list with secret keys to the Box.

Keys are stored in a database together with an identity derived from that key.
This identity derivation function can be a cryptographically secure hash functions such as SHA-256, making it impossible to deduce the secret key from the identity.

Besides distributing firmware and keys, the backend creates a wireless network to which sensor nodes connect and send data.
Upon receiving a message from a sensor node, the backend extracts the sender's identity and uses it as a search index in the database.
Thereby, the secret key assigned to that particular sensor node can be found efficiently.
With the secret key, the backend validates the authenticity of messages received and decrypts them.

\subsection{Sensor Nodes}
\label{Subsec:SensorNodes}
After power-up, the sensor node has no runtime-firmware image.
Subsequently, it queries a network address, on which the Box listens and responds.
Upon receiving a request, if signed firmware images are used, the Box prepares an \gls{OTA} bootloader with a public key and supplies it to the sensor node.
Using the new \gls{OTA} bootloader, the sensor node requests a runtime firmware-image.
The Box customizes a firmware image by injecting a secret key and delivers it. 
The sensor node validates this firmware image and writes it into its \texttt{Bootloadable} memory region and reboots, loading this image.
It contains functions to read sensors, the cryptographic key used for encrypting and authenticating messages as well as the address and network credentials of the backend.
Thus, the sensor node can immediately start reading sensors and transmit those readings to the backend securely.
Every sensor node derives an identifier from its secret key using the same derivation function as the backend and appends this identifier (or a digest from it) to messages.
This aids the backend in finding the secret key assigned to that particular sensor node.

%
Provisioning sensor nodes with the Box neither requires additional transceivers, nor I/O such as buttons or a display as it would be required for many \gls{LLC} approaches.
In the runtime-firmware, we do not need functions for interfacing this additional hardware, neither do we need a key exchange protocol.
We merely require the sensor nodes to have an \gls{OTA} bootloader and to be able to communicate with a transmission protocol capable of sending data from the backend to the sensor node.
This bootloader	can utilize a default network address to connect to.
We do not require customization of the bootloader.

As we use \gls{OTA} updates to program the sensor node, debug interfaces can be locked.
This aids in protecting the firmware from being read and keys from being extracted.
Using the Box, every sensor node has an individual secret key, which reduces damage in case a single sensor node gets captured and its firmware extracted.

We can leverage the support for multiple transmission protocols offered by modern \glspl{SOC}.
For example, the sensor node can acquire the firmware \gls{OTA} using WiFi for high-speed download.
If the sensor node should be used in a \gls{BAN} \cite{Dietz2018, Law2011}, the runtime-firmware image could contain a Bluetooth stack.
If long-range transmissions are needed, LoRAWAN could be used.
If the sensor node is assigned a different task, it can be brought back into \texttt{Bootloader} mode and then supplied with a new firmware image using the Box.
Hence, our approach permits convenient switching to the transmission protocol being the optimal choice for a given task.

%% file: Implementation.tex
\section{Implementation}
\label{Sec:Implementation}

\subsection{Box}

\begin{figure}
	\centering
	\includegraphics[width=0.9\columnwidth]{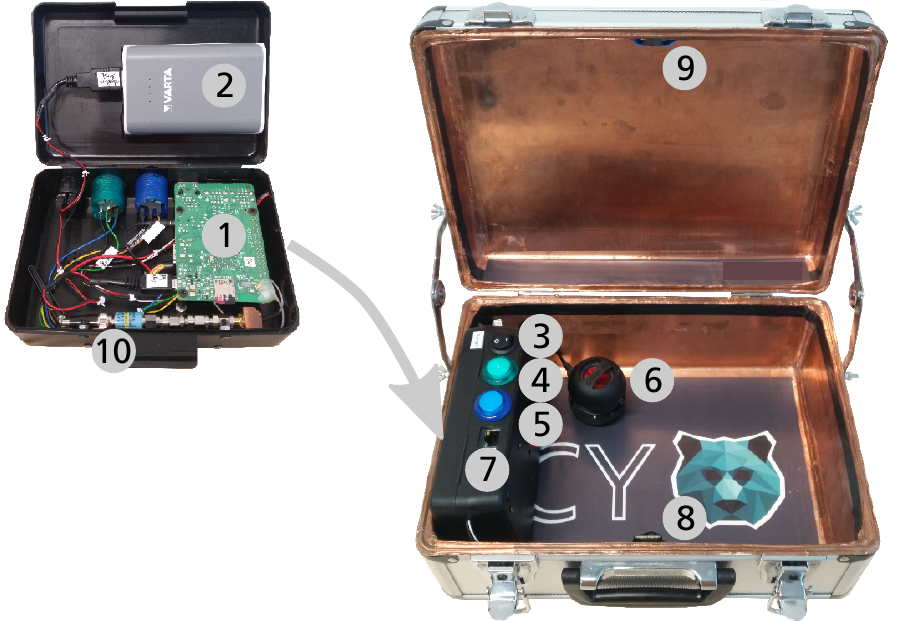}
	\caption{The \textit{Box} prototype}
	\label{Fig:Prototype}
\end{figure}

The Box is made up of a COTS lockable aluminum case spaced 320\,mm $\cdot$ 230\,mm $ \cdot$ 150\,mm.
We covered the interior of the case with copper-sheets and -tape and metered the attenuation provided by our DIY shielded Box to be at least 40\,dB.
This low-cost design permits creating a box for less than 100\,Euro, permitting widespread use.
If higher levels of shielding are desired, a commercial box should be used.

Figure \ref{Fig:Prototype} shows our prototype implementation of the Box.
Inside the Box, a \textit{Raspberry Pi Model 3} (Pi) \textcircled{1}, a battery pack \textcircled{2}, a power button \textcircled{3}, the \textit{deploy} \textcircled{4} and the \textit{acquire} pushbutton \textcircled{5} with built-in green and blue LEDs and a speaker \textcircled{6} are attached.
Using a tailored enclosure, only the buttons, the speaker and the Ethernet interface \textcircled{7}, which connects the Pi and the backend, are exposed to the user.
This prevents faulty operation by limiting the possible interactions.
The user is instructed about the current state and how to proceed by spoken output over the speaker.
Whenever the user is audibly instructed to press a particular button, the built-in LED of this button is flashed.

We monitor the state of the Box (e.g., whether its opened or closed) with a Hall Sensor \textcircled{8} connected to a GPIO pin and hooked to an interrupt handler. 
The magnet \textcircled{9}, which activates the Hall sensor, is attached to the lid.
All invalid transitions (e.g., open lid during firmware distribution) result in a secure fail state that deletes all secret data in the memory of the Box, as we must assume, that the key exchange has been compromised due to the absence of \gls{EM} shielding.

The Pi acquires firmware images and keys from the backend using Ethernet.
It can use its built-in WiFi and Bluetooth transceivers to communicate with sensor nodes.
To support other transmission protocols, additional transceivers can be attached via \gls{USB}.
In our tests we used the built-in WiFi interface of the Pi.
When deploying firmware (thus the Box being in closed state), the Raspberry Pi creates a WiFi \gls{AP}.
Firmware requests sent to the \gls{AP} are handled by a webserver, which in return sends the firmware patched with individual keys to the nodes.
Multiple requests can be handled simultaneously, thus, several sensor nodes can be served in parallel.

We decided to use the more powerful backend to build the images and transfer them to the Box rather than compiling in the Box.
Patching of firmware images is done by opening the image and replacing the key placeholder
with an individual key.
Subsequently, the Box signs firmware images.
By doing so, we are able to decrease the time needed in the field when deploying firmware images to nodes, as patching with keys and signing is much faster than compiling.
%

Security against eavesdropping is achieved by combining the \gls{EM} shielding of the Box with hardware signal attenuation.
For that, we replaced the Pi's built-in WiFi antenna with an SMA connector.
By chaining analog SMA-attenuators before the antenna \textcircled{10}, the transmission is reduced to the minimum power level at which the sensor nodes are still able to receive packets (about -90\,dBm for our sensor nodes).
This renders the external jammer needed by previous works obsolete.

\subsection{Backend}
\begin{figure}
	\centering
	\includegraphics[width=1\columnwidth]{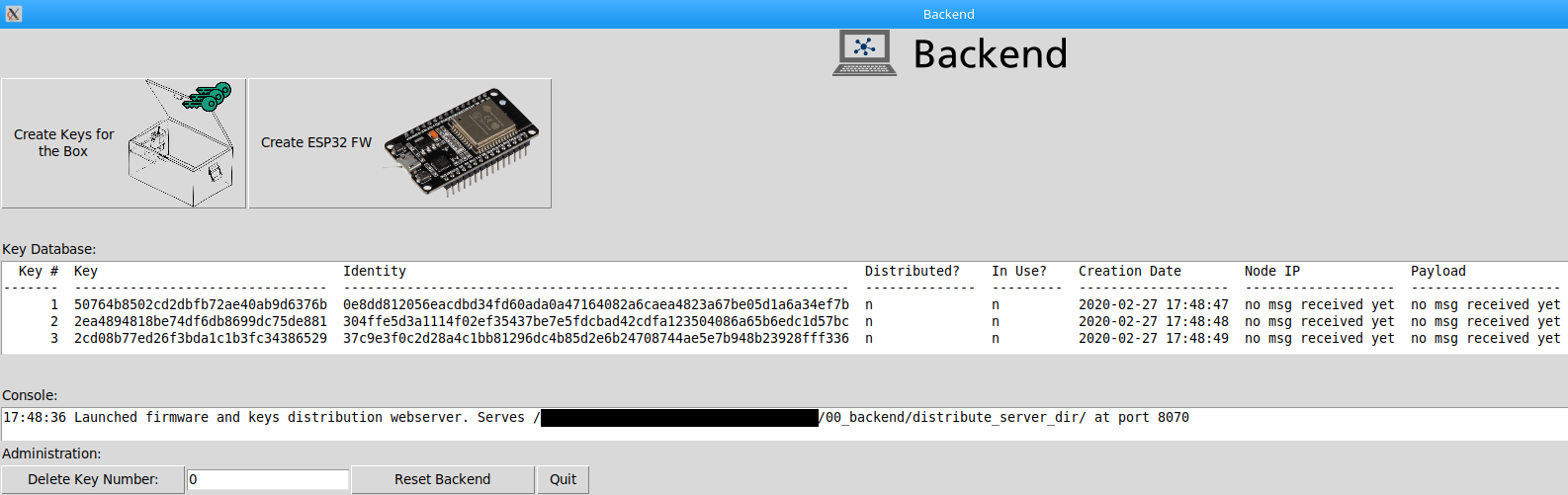}
	\caption{The backend \gls{GUI}}
	\label{Fig:GUI}
\end{figure}
Our backend is written in \texttt{Python3} and runs on an \gls{IPC}.
The \gls{GUI} built with the \texttt{tkinter} package is shown in Figure \ref{Fig:GUI}.
It has buttons for creating and deleting firmware images and secret keys.
The process of creating keys can also be performed automatically.

Communication between backend and Box uses \gls{HTTP} over Ethernet, however we plan to use HTTPS in the future.
To initiate the sensor node setup process, the user plugs the Box into the \gls{IPC}.
The backend runs a \gls{DHCP} server, which supplies the Box with an IP address.
It also runs a webserver, from which the Box requests a firmware image and secret keys up to a preset number.

For receiving sensor node readings, a WiFi \gls{AP} is created by the backend.
The sensor node knows the credentials of the access point as well as the address of the webserver, to which sensor readings shall be sent, as they have been included in the firmware images delivered to the sensor node by the Box.
Thus, immediately after setup, sensor nodes search for the backend WiFi network, join it and post sensor readings to the webserver.

\subsection{Sensor Nodes}
\label{subsection:sensornodes}

We implemented the sensor nodes using \textit{PYCOM LoPy4} development boards (based on the ESP32 \gls{MCU}) and the \textit{Pysense} sensor shield.
These nodes support four wireless interfaces: WiFi, \gls{BT}, LoRAWAN and SIGFOX. 
Further, they are capable of performing \gls{OTA} firmware updates over WiFi and LoRAWAN out of the box.
We used \gls{OTA} over WiFi with the default SSID, WiFi passphrase and webserver address set by the manufacturer.
The Box uses those defaults to launch an access point and a webserver, at which it serves firmware updates.
By using this setup, we do not have to adjust the \gls{OTA} bootloader, as the Box adapts to the sensor node and not the other way round.

%% file: Evaluation.tex
\section{Usability Study}
\label{Sec:Evaluation}


\paragraph{Purpose and Scope:} 
The aim of this usability study was to compare the speed and user-friendliness of setting up a \gls{WSN} using wired provisioning compared to our Box.
In the following, those two approaches are called \textit{Wired} and \textit{Box}.
The task for the subjects was to supply firmware to sensor nodes using both approaches.
Wired can be considered the traditional approach for small volumes of sensor nodes.
%
It is automated to some extend using a script, which takes care of creating and injecting a unique key into the sensor node firmware image and writing the image to the node via \gls{UART}.
%
%
%

We investigate the following hypotheses:
\begin{enumerate}
	\item Using the Box to provision sensor nodes needs less time than using wired provisioning.
	The advantage in speed increases, the more sensor nodes are to be provisioned.
	%
	\item Using the Box is intuitive and does not permit the user to make security-critical errors.
\end{enumerate}

\paragraph{Experimental Design and Setup:}
The two approaches \textit{Box} and \textit{Wired} are compared within-subjects, i.e.~all participants performed both approaches.
We used LoPy4 sensor nodes and a laptop running \texttt{Ubuntu 18.04}. 
To interface the laptop with the LoPy, we used \texttt{esptool.py}, a command-line tool, for which we created a clickable desktop icon for easier use.

%
%
Subjects were given the task to set up a \gls{WSN} by programming sensor nodes.
They were further instructed, that upon issuing a particular code word, the experiment supervisor would show up and assist them.
The code word prevented accidental or inexplicit calls for help.

\paragraph{Sample:} We recruited a total of $N=\numberof$ subjects.
21 of them were male, 10 female.
The age of the subjects ranged between 23 and 53 years.
$N_{1} = \numberofnerds$ subjects were from computer science and electrical engineering study programs.
For those, we assumed a certain familiarity with networking, embedded systems and security.
Hereinafter, they will be called \textit{experts}.
To test whether the Box approach is suitable for users with no technical background, we recruited additional $N_{2} = \numberofothers$ participants.
Five of those had a academic degree we considered unrelated to the task, while six did not have an academic degree at all.
%
%
Hereinafter, they will be called \textit{novices}.

\paragraph{Procedure:} Prior to execution, the supervisor explained the test scenario, described the tasks to be performed and explained his own role as technical support.
Next, subjects were given a printed instruction sheet in their native language.
They were asked to read aloud the instruction sheet and ask questions on the instructions.
There were no questions raised in the final study, as we had evaluated the instruction sheet in a pilot study with six subjects and adjusted it to answer all questions raised there.

Subjects were provided with a switched-on laptop, its desktop clearly showing the executable icon for initiating wired programming, and four LoPy4 sensor nodes.
They were asked to supply firmware to those four sensor nodes using \textit{Wired}.
For that, they were instructed to connect the sensor node to the computer used for programming, hold a pushbutton to enter the bootloader mode, execute the script and finally unplug the node.

To evaluate the Box approach, subjects were given the Box prototype and four LoPY4 sensor nodes, all switched on prepared with an \gls{OTA} over WiFi-capable bootloader.
The Box already held a firmware image and a sufficient number of secret keys in its memory.
This setup resembles a scenario in which the shift manager had supplied keys to the Box at the beginning of the shift. 
A normal worker provisioning the sensor nodes would not be permitted to interact with the Backend system.
Subjects were instructed to place all four nodes together in the Box, press the deploy-button and close the Box.
Then they were supposed to wait, while the Box programmed sensor nodes, until the Box instructed them to open it and remove the sensor nodes.

To control for learning effects, e.g., embedded device handling, after each participant, the order of \textit{Wired} and \textit{Box} was randomized.
Further, this accounts for opinions from one approach influencing the opinion on the other approach. 
%
%
After subjects had completed the firmware distribution process, we asked them for qualitative feedback on their experience with the Box using the 10-item Likert scale \gls{SUS}, which is widely used \cite{brooke1996sus}.

\paragraph{Measurement:}
We measured the \textit{time needed} to provision sensor nodes from the moment the subject consciously stated "start" until all sensor nodes were provisioned successfully.
We counted the \textit{number of hesitations}, \textit{errors} and \textit{calls for support} made.
Hereby, \textit{hesitation} denotes a time span of at least three seconds, in which the subject did not proceed.
Hesitations are not necessarily related to an \textit{error}, which denotes doing something different from the instruction sheet.
Those metrics map to the usability attributes \textit{learnability, attractiveness, user-friendlyness, low cost to operate} and \textit{security interaction} according to ISO 9241-110:2006 \cite{ISOUsability}.

\subsection{Results}
Due to the small sample size data were only analyzed descriptively.
Results are summarized in Table \ref{Tab:studyresults}.

\textit{Time Needed:}
Using \textit{Wired}, both experts and the novices took around 42\,s to program the first sensor node.
For sensor node four, it took them an average of 23\,s.
One run of the Box with four sensor nodes inside took the experts an average of 30.9\,s and the novices 29.7\,s.

\begin{table}
	\caption{Study results. Times given for wired deployment are per node, for the Box for one run with four nodes.}
	\label{Tab:studyresults}
	\begin{tabularx}{\columnwidth}{ p{0.2cm} p{2.4cm} | X  X X X | X }
		& & \textbf{Wired} & &  & & \textbf{Box} \\
		\hline
		\multirow{5}{*}{\rotatebox{90}{\textbf{Novices}}} & \multirow{2}{*}{Time needed [s]:} & \#1 & \#2 & \#3 & \#4 & total \\
		& & 42.3 & 35.3 & 26.3 & 24.2 & 29.7\\
		& Hesitations: & 2 & & & & 1 \\
		& Errors: & 3 & & & & 0\\
		& Support calls: & 0 & & & & 0 \\
		\hline
		\multirow{5}{*}{\rotatebox{90}{\textbf{Experts}}} & \multirow{2}{*}{Time needed [s]:} & \#1 & \#2 & \#3 & \#4 & total \\ 
		& & 42.9 & 29.9 & 27.2 & 22.6 & 30.9\\
		& Hesitations: & 5 &  & & & 2\\
		& Errors: & 4 &  & & & 1 \\
		& Support calls: & 0 & & & & 2 \\
		
	\end{tabularx}
\end{table}

\textit{Hesitations and errors:}
With \textit{Wired}, the novices hesitated twice and made three errors.
All of them could be attributed to the subject struggling with the button to enter the bootloader mode.
With the Box, one hesitation and zero errors were counted.
The subject thought only a single node was to be placed in the Box.
After re-reading the instruction sheet, the subject placed the remaining sensor nodes in the Box and proceeded.

The experts hesitated five times and made four errors using wired distribution, while there were two hesitations and one error using the Box.
All errors and hesitations during wired distribution were related to the bootloader mode button, e.g., the button not being pressed, searching the button at the laptop screen rather at the sensor node or pressing and releasing the button immediately.
The hesitations caused by the Box are related to the subjects not having understood the audio feedback and hence, they were not sure whether the programming process had finished.
The error was caused by the subject turning off the Box, despite the power switch being in the clearly marked "on" position and the green deploy button flashing.

\textit{Number of support calls:}
The novices requested the support zero times for both approaches, while the experts asked two times using the Box.
Both occurrences were caused by the Box prototype not responding to the deploy button press.

\textit{\gls{SUS}:}
Results from the \gls{SUS} questionnaire are shown in Table \ref{Tab:susresults}.
Subjects gave overall good grades.
Differences between experts and novices are negligible. 
\begin{table}
	\caption{\gls{SUS} results}
	\label{Tab:susresults}
	\begin{tabularx}{\columnwidth}{l X X X X X X X X X X}
		Question \# &1&2&3&4&5&6&7&8&9&10 \\
		\hline
		Experts: &4.9&1.2&5.0&1.3&4.8&1.1&4.9&1.1&4.3&1.2\\
		Novices: &4.8 & 1.0 & 5.0 & 1.6 & 4.8 & 1.2 & 5.0 & 1.0 & 4.9 & 1.1 \\
	\end{tabularx}
\end{table}

\textit{Qualitative feedback:}
Subjects described the Box as very practical and intuitive to operate.
They liked the gain in speed over the wired approach and that the Box supersedes cable handling for placing devices and operating the Box lid.
The voice instructions were perceived helpful and could be understood easily.
Subjects felt reminded of automated external defibrillators, which also give step-by-step instructions.
As the Box informed the user that firmware exchange was finished and the Box could be opened, subjects were able to judge the success of deployment reliably.

\subsection{Discussion}
%
As expected, the Box outperforms wired distribution in terms of speed and scalability.
Two aspects contribute to the total time needed for one run:
Firstly, time needed for devices to communicate, download the firmware and write the new firmware image to flash memory.
Secondly, the time needed for device handling.
Concerning the former, the gain in speed stems from \gls{OTA} programming inside the Box taking place in parallel, while the wired approach is limited to programming one sensor node at a time.
For device handling time, independently from previous experience, both groups became quicker while handling the sensor nodes in the wired approach. 
Assuming the fastest time needed for wired programming of one sensor node, which is about 23\,s, we extrapolate that provisioning two nodes takes about 46\,s.
Using the Box to set up \textit{four} sensor nodes took both groups about 30\,s.
Thus, we can conclude, that the Box is faster than wired provisioning, as soon as more than one sensor node is to be programmed. 
Using the Box, device handling times varied, because some subjects arranged sensor nodes neatly inside the Box, while others just took the bunch of four and dropped them randomly.
While sensor nodes can be successfully provisioned at any position inside the Box, as stated in Section \ref{Subsec:Box}, the operator should be aware, how many sensor nodes have been placed inside.
Hence, it seems useful to structure the interior of the Box such that one sensor node resides at one position, e.g., by adding foam-cutouts.

The \gls{SUS} and the qualitative feedback show, that subjects consider the Box intuitive, easy to use and well integrated.
Especially subjects, who had received industrial training, liked, that operating the Box requires only simple manual actions, which they felt confident doing.
They also liked, that the Box hides away the electrical and software aspects, which are exposed in wired programming.
Subjects stated, that this reduced their worries of breaking something.

One error was made, as the subject turned the already switched Box off.
Support was called twice, as the Box prototype did not respond to the button press.
Two subjects from the experts group wished to be able to understand better, what is happening internally.
They asked for more fine-grained information, e.g., with which sensor node the Box is communicating at the moment.
To account for that, the Box should provide more precise feedback more often by playing repeated audio feedback without the user performing an action.
Alternatively, a help button could be added, which, upon being pressed, tells about the current state and how to proceed.
Overall, all subjects felt more confident and expressed higher satisfaction using the Box compared to wired programming.

\paragraph{Outlook:}
In this study, we did not induce arbitrary errors to investigate, how the subjects recover from those.
This is not realistic in our industrial scenario: In case of an error, a subject will always call the supervisor and ask for help.
Here, we favored a more realistic experiment over trying to evaluate edge cases.
As future work, we are going to evaluate the Box in a industrial production environment to learn, how the subjects cope with the Box in harsh environments.

%
%

%% file: Conclusion.tex
\section{Conclusion}
\label{Sec:Conclusion}
\glsresetall
We have designed and implemented the \textit{Box}, an intelligent \gls{FC}, which can supply firmware images and secret keys to large numbers of sensor nodes in a user-friendly manner using \gls{OTA} updates.
Previously proposed approaches are based on the assumption, that the firmware is "just there", omitting, \textit{how} the distribution of firmware is done.
In contrast, we demonstrate a complete workflow for both firmware deployment and the exchange of device-individual secret keys leveraging \gls{OTA} firmware transmission protected by the \gls{FC}.
\gls{OTA} requires a wireless transmission protocol, which permits transmissions from an update server to a sensor node.
As this is the case for transmission protocols commonly used in \glspl{WSN}, the Box can be used with most \gls{COTS} sensor node platforms.

The proposed workflow shifts security relevant tasks from the operator to the Box, thus minimizing room for human error. 
Manual actions are reduced to opening and closing the box and placing and removing sensor nodes.
The Box instructs the operator via spoken commands.
Our usability study shows, that both technically adept and non-tech-savvy operators are able to use the Box intuitively and can deploy firmware faster compared to the classic wired approach.